# Atomic Scale Insights Into The Mechanical Characteristics of Monolayer 1T-Titanium Disulphide: A Molecular Dynamics Study

Tanmay Sarkar Akash,[a] Rafsan A.S.I. Subad,[a] Pritom Bose,[a] and Md Mahbubul Islam*[b]

[a]Department of Mechanical Engineering, Bangladesh University of Engineering and Technology, Dhaka-1000, Bangladesh

[b]Department of Mechanical Engineering, Wayne State University, 5050 Anthony Wayne Drive, Detroit, MI- 48202, USA

*Corresponding Author Tel.: 313-577-3885; E-mail address: mahbub.islam@wayne.edu

**ABSTRACT**

In this work, we report on the mechanical responses and fracture behavior of pristine and defected monolayer 1T-Titanium Disulfide using classical molecular dynamics simulation. We investigated the effect of temperature, strain rate and defect ratio on the uniaxial tensile properties in both armchair and zigzag direction. We found that monolayer $TiS_2$ shows isotropic uniaxial tensile properties except for failure strain which is greater in zigzag direction than armchair direction. We also observed a negative correlation of ultimate tensile strength, failure strain and young's modulus with temperature and defect ratio. Results depicts that strain rate has no effect on the young's modulus of monolayer $TiS_2$ but higher strain rate results in higher ultimate tensile strength and failure strain.

## 1. INTRODUCTION

Transition Metal Dichalcogenides (TMDs) are being studied extensively nowadays due to their distinctive properties arising from their two dimensional (2D) structures,[1–4] Graphene, as the first discovered 2D material, with a honeycomb lattice structure[5] having extraordinary electric and thermal properties,[6,7] lacks direct bandgap which hinders its applications in optoelectronic field-effect transistors (FETs).[8] Here, TMDs can act as a prospective substitute for graphene owing to their 1–2 eV direct bandgap [9,10] and a wide range of applications [11–14] for instance energy storage devices, electronic devices, sensors, and so on.

A TMD consists of a sandwiched structure having strongly covalent bonded one hexagonal layer of metal atoms (M) and two layers of chalcogen atoms (X) with a common formula of $MX_2$.[15] Generally, most of these materials seem to have either in the trigonally coordinated H phase or the octahedrally coordinated T phase, and very few of them are stable in both T and H phases.

Numerous methods, such as mechanical cleavage,[16] compressible flow exfoliation,[17] liquid exfoliation,[18–20] and chemical vapor deposition (CVD),[21–24] have already been used to prepare ultra-thin TMDCs. Lately, an electrochemical Li-intercalation method is also established to prepare single-layer metal disulfide, such as $MoS_2$, $WS_2$, $TiS_2$, and $TaS_2$,[25] and few-layer metal selenide, such as $WSe_2$, $NbSe_2$, and $Sb_2Se_3$.[26] Their impressive electronic,[27] optical,[28] structural,[29] and transport[30] properties along with the corresponding large surface area have allowed them to come out as the prospective candidates for not only in the field of optoelectronics,[31] but also in the energy scavenging that includes energy conversion,[32] storage,[33] Hydrogen Evolution Reactor (HER),[34] and catalysis.[35,36]

Among TMDs, Titanium Disulfide (TiS$_2$) is the lightest in weight, highly stable and a low cost material.[37] TiS$_2$ promises quite a number of possible applications which are being examined as cathode material in magnesium-ion batteries,[38] as (lithium-intercalated) high-energy storage systems, and for catalysis,[36] etc. It is also one of those TMDs which can exhibit the properties of both metal (superconductor) and semiconductor, and, therefore, has gained substantial attention in recent days. It has been recognized that the octahedral atomic structure (1T) is the only stable state for monolayer TiS$_2$ crystalline.[37] There are reports showing that the electronic structure, and also the transport properties in TiS$_2$ can be altered by means of external pressure or applying mechanical strain.[15,27,36,39] This resembles the necessity of the investigation on the mechanical properties of monolayer TiS$_2$ sheet.

However, in many situations, defects in the single layer sheet of TMDs are hard to elude. Several defects in 2D sheets[40–42] are predominant in their chemical growth processes. Radiation damage in the applications of TMDs as nano-catalysts and dry lubricants obviates defect formation in the sample and severe structural damage might be happened if they are not cautiously fabricated and maintained. Often defects are purposefully imposed upon the nanostructures to enhance the preferred properties. Hence, it is also indispensable to inspect the mechanical properties defected single layer TiS$_2$ (SLTiS$_2$) sheet to predict and prevent the premature mechanical failure.

In the recent past, numerous studies have been conducted based on first principle calculations[27,37] to inspect the electrical, structural, and optical properties of monolayer TiS$_2$ sheet. However, Density Functional Theory (DFT) calculation is only used to simulate a few number of atoms. It is too much time consuming otherwise. As a result, it is nearly impossible to find some properties where it requires a lot of atoms to simulate such as the effect of defect concentration, and to demonstrate the understanding of fracture mechanism of a TiS$_2$ nanosheet using DFT calculation. Nevertheless, to find out the mechanical properties of single to few layer TiS$_2$, the amount of study is still wanting. Moreover, there is hardly any study for determining the mechanical properties of SLTiS$_2$ regarding molecular dynamics (MD) technique.

Therefore, in this study we have demonstrated the mechanical properties of pristine TiS$_2$ sample varying temperatures from 10K and 600K for a constant strain rate $10^9$ s$^{-1}$. Then, we varied strain rate ranging from $10^8$ s$^{-1}$ to $10^{10}$ s$^{-1}$ for a fixed temperature. We also investigated the consequence of different types of defects and the effect of defect ratios varying from 0% to 5%. Finally, we also explain the fracture mechanism in details.

## 2. METHODOLOGY

We created 20nm x 20nm monolayer 1T-TiS$_2$ sheets using a MATLAB[43] script. The lattice constant of 1T-TiS$_2$ used here is a=b=3.32Å[44]. Stresses are calculated without including the actual thickness of the 2D sheet. In order to study the effect of different defects[45] on the mechanical behavior of nano TiS$_2$ sheet, several types of defected (Titanium Vacancy, Sulfur Vacancy, Frenkel, Antisite STi, Antisite TiS, Three neighboring S vacancies under a S, Three neighboring S vacancies under a Ti, Three S vacancies divided by a TiS6 octahedron, Ti3S vacancy, TiS3 vacancy, TiS6 vacancy) sheets were created by deleting atoms or by manipulating their positions. We further varied the vacancy concentration from 1% to 5% to discover their effect on mechanical properties. For a statistically sound analysis, we used four different sheets containing vacancies at random positions and constructed stress-strain relationship (see supplementary Figure S3). MD simulation was performed to determine the tensile mechanical properties and to study the fracture mechanism of monolayer TiS$_2$. All simulations were performed using LAMMPS[46] simulation package. Periodic boundary condition was kept in the planar directions (x and y) in order to avoid edge effects while z-direction was kept non-periodic. A time step of 1 femto-second was used. At first we derived an energy-minimized structure using conjugate gradient (CG) scheme. The structure was then relaxed at finite temperature using an NPT ensemble. Pressure and temperature damping constants were kept 1, and 0.1 ps respectively in both the planar directions. Stress-strain relationship is determined by deforming the simulation box uniaxially and calculating the average stress over the structure. Atomic stress was calculated on the basis of the definition of virial stress. Virial stress components are calculated using the following relation:

$$\sigma_{virial} = \frac{1}{\Omega} \sum_i \left( -m_i \dot{u}_i \otimes \dot{u}_i + \frac{1}{2} \sum_{j \neq i} r_{ij} \otimes f_{ij} \right) \quad (1)$$

where $\Omega$ is the volume occupying the atoms, $\otimes$ is used to imply the cross product, the mass of atom i is represented by $m_i$, the position vector of the atom is $r_{ij}$, $\dot{u}_i$ is the time derivative and signifies the displacement of an atom from a reference position, and $f_{ij}$ is the interatomic force applied on atom i by atom j. Engineering strain is entitled by the term strain and calculated as:

$$\varepsilon = \frac{l - l_0}{l_0} \quad (2)$$

Here, undeformed length of the box is denoted by $l_0$, and $l$ is the instantaneous length.

In order to address the interatomic interactions, a recently establised Stillinger–Weber (SW) potential by *Jiang et al.*[47] is employed. The SW potential has of a two body term and a three body term relating the bond stretching and bond breaking, accordingly. The mathematical expressions are as follows:

$$\Phi = \sum_{i<j} V_2 + \sum_{i>j<k} V_3 \quad (3)$$

$$V_2 = A e^{\left[\frac{\rho}{r - r_{max}}\right]} \left(\frac{B}{r^4} - 1\right), \quad (4)$$

$$V_3 = K\varepsilon e^{\frac{\rho_1}{r_{ij} - r_{max\,ij}} - \frac{\rho_2}{r_{ik} - r_{max\,ik}}} (\cos\theta - \cos\theta_0)^2 \quad (5)$$

Here $V_2$ and $V_3$ the two body bond stretching and angle bending terms accordingly. The terms $r_{max}$, $r_{max\,ij}$, $r_{max\,ik}$ are cutoffs and $\theta_0$ is the angle between two bonds at equilibrium configuration. A and K are energy related parameters that are based on VFF model. B, $\rho$, $\rho_1$, and $\rho_2$ are other parameters that are fitted coefficients. These parameters and their corresponding value can be found in ref.[47].

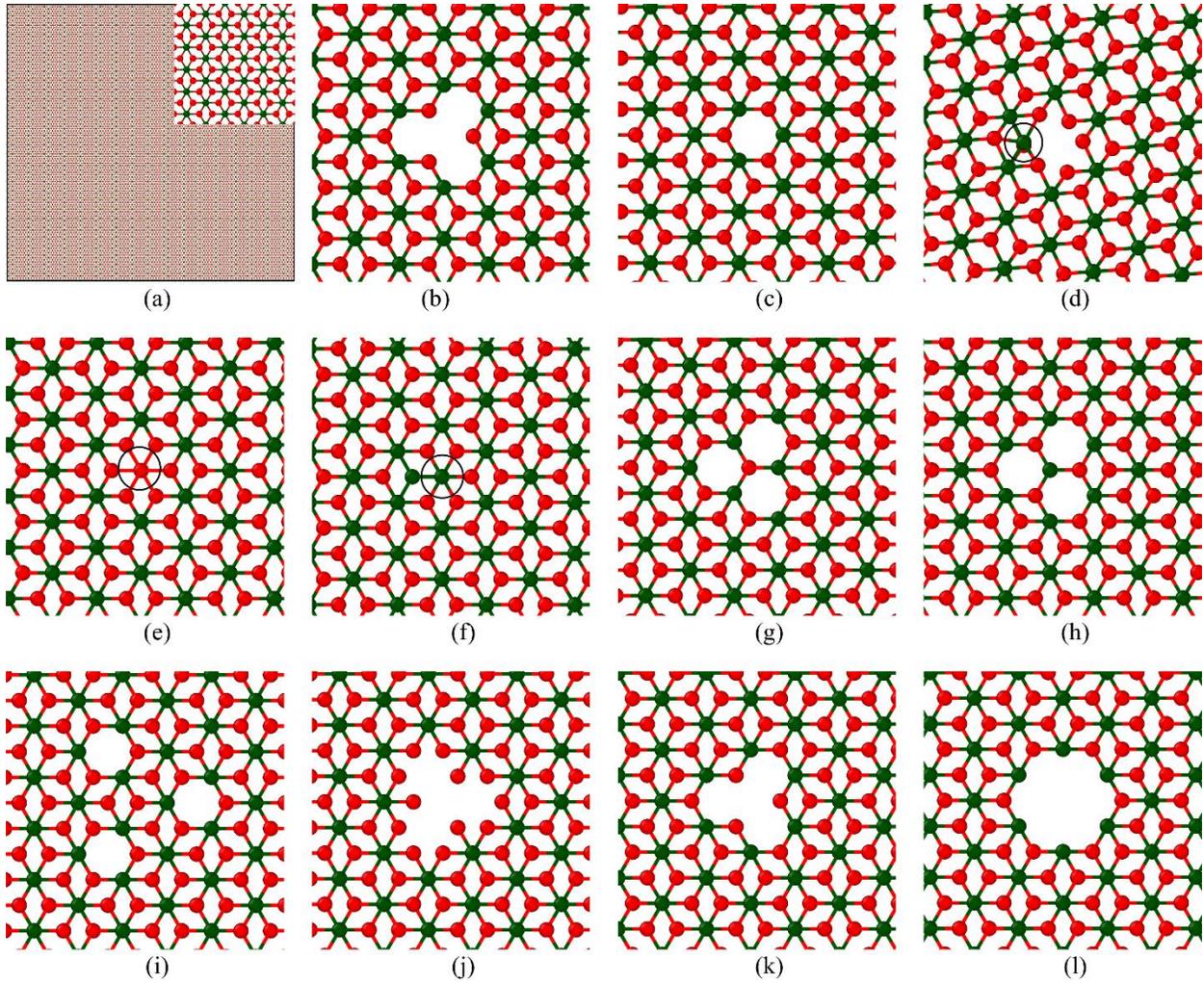

Figure 1: Atomic structure of (a) pristine TiS$_2$ monolayer (b) Ti vacancy (c) S vacancy (d) Ti Frenkel defect (e) Antisite STi defect (f) Antisite TiS defect (g) Three neighboring S atom vacancies under a S atom (h) Three neighboring S atom vacancies under a Ti atom (i) Three S atom vacancies divided by a TiS6 octahedron (j) TiS3-type vacancy (k) Ti3S-type vacancy (l) TiS6-type vacancy. Atom color: S, red; Ti, green.

## 3. METHOD VALIDATION

Stress-strain co-relation has been constructed (see supplementary Figure S1) for pristine monolayer TiS$_2$ nanosheet at 10K temperature to validate the method used in the study. The calculated fracture strength, fracture starin, Poisson's ratio and Young's modulus are then compared with literature.[47,48] The comparison among the results is presented in Table 1. Here, although it seems that the Young's modulus is slightly lower than the first principle calculation[48], it can be explained by the fact that the potential utilized here considers only short range interaction which generally results in underestimation of Young's modulus by 10%.[47] Therefore, it can be ensured that the projected results in the present study are in good harmony with the literature.

Table 1: Comparison of our calculation with the existing literature

| Mechanical Properties | This Study | | Previous Study[47] | | Previous Study[48] |
|---|---|---|---|---|---|
| | armchair loading | zigzag loading | armchair loading | zigzag loading | |
| Young's modulus (N/m) | 74.7 | 74.2 | 75 | 74.6 | 85 |
| Poisson's ratio | 0.20 | 0.20 | 0.20 | 0.20 | 0.20 |
| Fracture strength (N/m) | 10.8 | 10.4 | 10.6 | 10.2 | |
| Fracture strain (%) | 25 | 29 | 25.1 | 28.8 | |

## 4. RESULTS AND DISCUSSIONS

### 4.1 Effect of Chirality

In case of 2D materials, chirality drives a crucial role in defining the structural properties.[49,50] In order to investigate the effect, we applied uniaxial tensile stress on $SLTiS_2$ nanosheet keeping a constant temperature and strain rate. The inquiry reveals that there is a little dependence on the material properties over chirality (see supplementary Figure S1). Poisson's ratio is found to be 0.20 in both direction while Young's modulus, fracture strength, and fracture strain are observed to be 74.7 N/m, 10.8 N/m, 25% and 74.2 N/m, 10.4 N/m, 29% for armchair and zigzag loading direction accordingly. Because of the three fold rotational symmetry of $1T-TiS_2$ crystal lattice, the Young's modulus and other second order elastic constants are nearly isotropic.

Again, in the event of armchair tension, there is a bond exactly along the loading direction which works in the load bearing and therefore, stress is increased extensively by increasing the strain along armchair direction. On the contrary, during zigzag tension, one bond is almost ($\pm 30°$) and another is precise perpendicular to the loading direction. Thus, the load bearing bond has to carry a fraction of the load exerted. Hence, the bond is less stretched than in case of armchair loading. This is the reason for the slightly higher tensile strength of $1T-TiS_2$ in armchair loading condition than the zigzag counterpart. Like InSe[51], the maximum difference in Young's modulus of 0.46N/m in between armchair and zigzag directional loading certifies the practical material isotropy of the monolayer $TiS_2$.

## 4.2 Temperature Effect

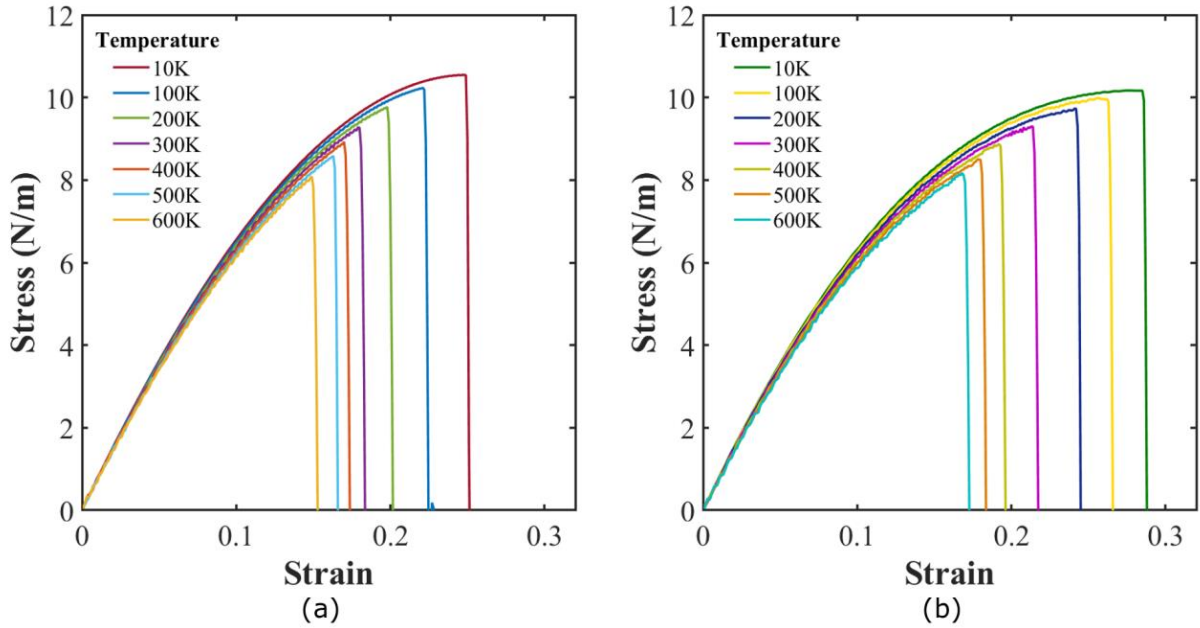

Figure 2: Temperature dependent stress-strain relationship of pristine TiS$_2$ sample having a fixed strain rate $10^9$ s$^{-1}$ along (a) armchair (b) zigzag direction

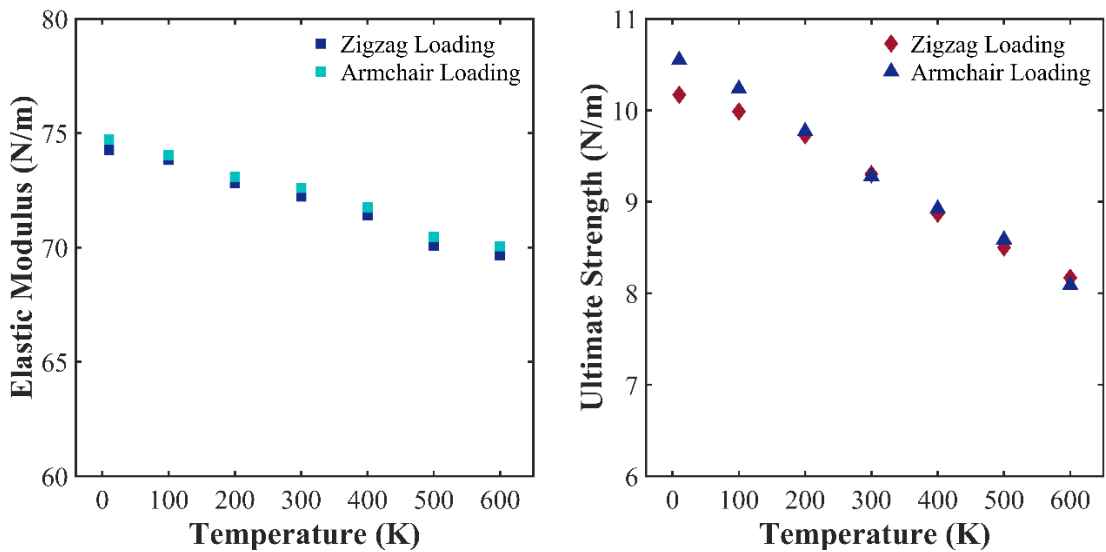

Figure 3: Variation of (a) elastic modulus and (b) ultimate strength of pristine TiS$_2$ with the change of temperature

Temperature is a governing issue for device performance which plays an ample impact on the electrical and mechanical properties of the material in devices. And, TMDs might face a threat of temperature due to the generation of Joule heat in case of real-life operations of electronic devices. Effects of temperature on the tensile strength and Young's modulus of a monolayer graphene sheet and multilayer $MoS_2$ are studied by Dewapriya et al,[52] and Ghobadi et al[53] respectively. Both of them disclosed that increased temperature results in the minimization of elastic constants, fracture strain, and fracture stress.

For both armchair and zigzag directions, the effect of temperature on the stress-strain relationship of single layer $TiS_2$ has been studied for a temperature range from 10K to 600K keeping a fixed strain rate of $10^9$ s$^{-1}$. It is obvious that increasing temperature speeds up the failure in all the cases, and fosters a significant reduction in fracture stress. It is evident from the Figure2 that for both directional loading conditions, the ultimate stress, elastic modulus, and failure strain reduces with the escalation of temperature. Since higher temperature encourages higher atomic movements which enhances the thermal vibrational instabilities. This not only makes bonds stretchier and the material softer but also enables the possibility of some bonds surpassing the critical bond length and instigating failure. Furthermore, the rise in temperature causes higher entropy in the material which emboldens crack propagation. Thus, material strength weakening takes place.

We calculated the elastic modulus of the material with the variations of temperature. Elastic modulus and the ultimate tensile stress of the material always declines as the temperature upsurges which has been depicted in Figure3 (a) and (b) accordingly. Elastic modulus and ultimate tensile strength show ~6% and ~22% reduction respectively from temperature ranging from 10K to 600K.

### 4.3 Strain rate sensitivity

The effect of change of strain rate on the monolayer $TiS_2$ sheet is demonstrated in the Figure4. $SLTiS_2$ sheet displays strain rate sensitive behavior. However, we discover that sensitivity to the applied strain rate is not as strong as other factors, for example, temperature and defect concentration. To find out the effect of strain rate, MD simulations have been carried out on $SLTiS_2$ nanosheet keeping a constant temperature 300K while varying strain rate from $10^8$ s$^{-1}$ to $10^{10}$ s$^{-1}$. Even though the strain rate range is quite higher relative to the real life ones, it is recurrent in MD simulations to qualitatively evaluate the response of strain rate at this order.[54,55] Ultimate tensile strength of $TiS_2$ nanosheet upturns almost ~5% with the rise in strain rate from $10^8$ s$^{-1}$ to $10^{10}$ s$^{-1}$. The reason can be explained by the fact that at higher strain rate, the response and relaxation time is so less that it causes less atomic thermal fluctuations to get over the energy barrier to break their bonds and cannot promote bond rearrangement, vacancy coalescence, and crack propagation, thus giving intensification to increase fracture stress and vice versa.[29,56,57] The fact is also sustained for $SLTiS_2$ sheet which has been shown in Figure4, illustrating higher fracture stress, and strain for higher strain rate.

Additionally, the responsiveness of fracture strength on strain rate is assessed in terms of the parameter m, strain rate sensitivity, calculated by the following equation[57]:

$$\sigma = C\dot{\varepsilon}^m \qquad\qquad 6$$

Here, $\dot{\varepsilon}$ is strain rate, m is the strain-rate sensitivity, and C is a constant. Equation can be presented in the logarithmic function as following:

$$\ln(\sigma) = \ln(C) + m\ln(\dot{\varepsilon}) \qquad\qquad 7$$

m can be calculated from the slope of the linear fitted data in Figure5. Evaluating the numerical values, we propose the equation for monolayer TiS$_2$ as:

In armchair direction:  y = .0086x + .891                                                      8

In zigzag direction:  y = .0137x + .847                                                        9

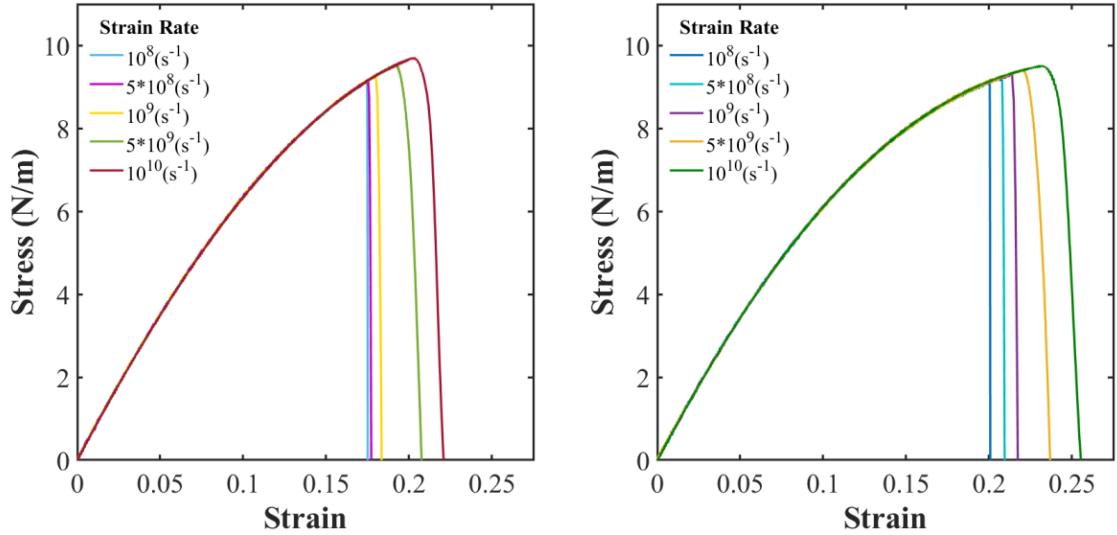

Figure 4: Stress-strain curve for strain rate dependency for TiS$_2$ loading along (a) armchair (b) zigzag direction

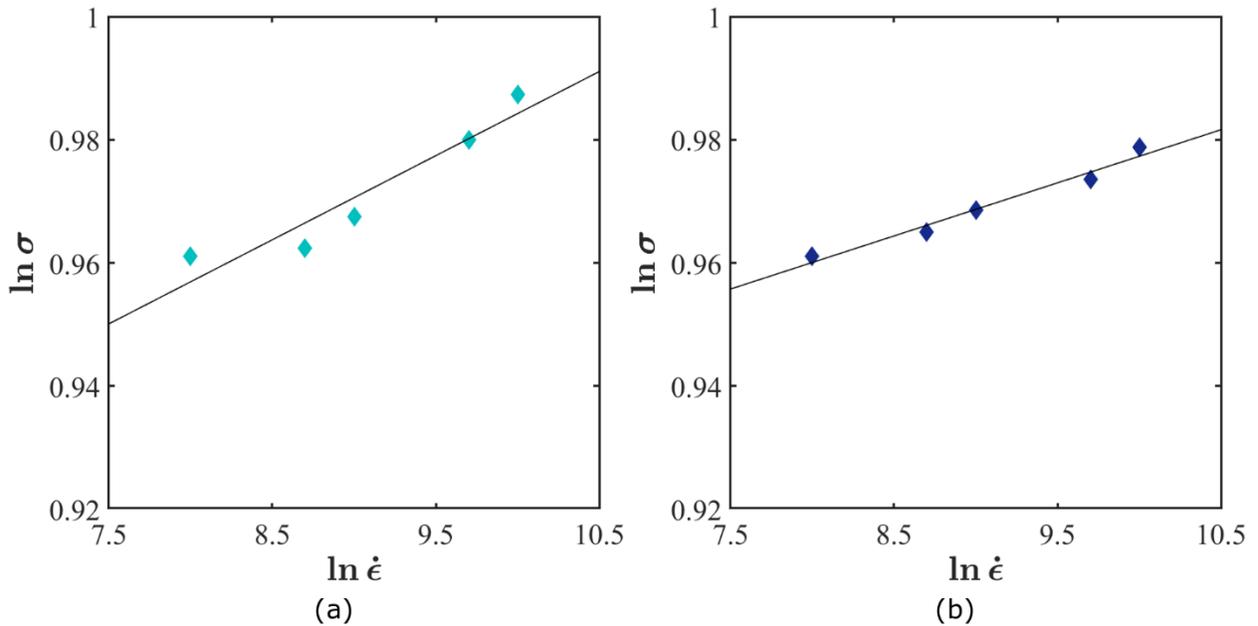

Figure 5: Logarithm variations of fracture strength with strain rate in (a) armchair (b) zigzag directions for pristine TiS$_2$

## 4.4 Effect of Defect

Fabricated $TiS_2$ inherently possesses many defects.[45] Furthermore, defects are sometimes intentionally introduced upon the nanostructures to engineer desired properties, exclusively electrical and optical properties and even to improve mechanical stability for some nanomaterials.[58,59] Presence of different atomic defects has extensive effect on the properties of 2D materials. Different kinds of defects can appear in 1T-$TiS_2$ sheet. Among them vacancy defects (either single or double point defects like Ti or S atom vacancy, Ti Frenkel defect or groups of point defects like three S vacancies divided by a TiS6 octahedron, three neighboring S vacancies lying under the S atom or under Ti atom, three Ti vacancies and an S vacancy etc) and antisite defects are considered in this study.[45,60]

To understand the effect of defects on monolayer $TiS_2$ nanosheet, stress-strain relationship has been constructed inducing different type of defects on the sheet while keeping the temperature and the strain rate constant. The stress-strain plot has been depicted in supplementary Figure S2. All kinds of defects substantially minimize the fracture strength and strain of the material for both armchair and zigzag direction. The structure fails before reaching the ultimate tensile strength and fracture strain of pristine 1T-$TiS_2$ sheet. Because significant stress concentration can occur around the vacancy defects which facilitates crack initiation and ultimately nucleation and propagation.

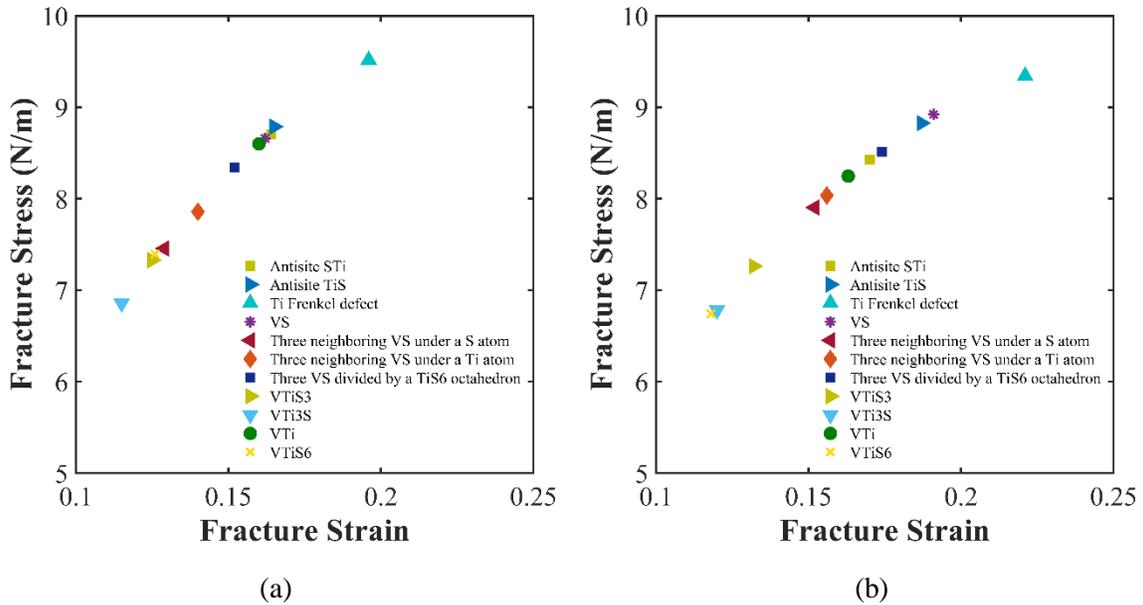

(a)          (b)

Figure 6: Change in fracture stress with the variations of defect types while loading along (a) armchair (b) zigzag direction

## 4.5 Effect of Point Defect Ratio

Increasing percentage of point defect ratio heavily impacts the elastic properties of the sheet. Figure 8 and 9 demonstrate that fracture strength, fracture strain and Young's modulus decrease almost 54.25%, 60% and 23.6% accordingly when the defect ratio increases from 0% to 5%.

Randomly distributed and increasing amount of defect concentration can considerably lessen the material integrity. It can be explained by the fact that the presence of vacancy defects at various positions of the sheet intensifies the concentrated stress at vacancy defect points and thus contributes to the increment of fracture nucleation points[61] which eventually breaks the structural integrity of $TiS_2$ monolayers, and facilitates premature failure as well.[62–64] Linear degradation in Young's modulus with the increase of defect density is also persistent with other studies for different 2D materials.[65,66] Assessing the numerical values we propose the linear relation for $SLTiS_2$ as:

For armchair tension:  y = -3.41x + 71.38            10

For zigzag tension:  y = -3.65x + 72.47             11

where, x and y represent Young's modulus and defect concentration respectively.

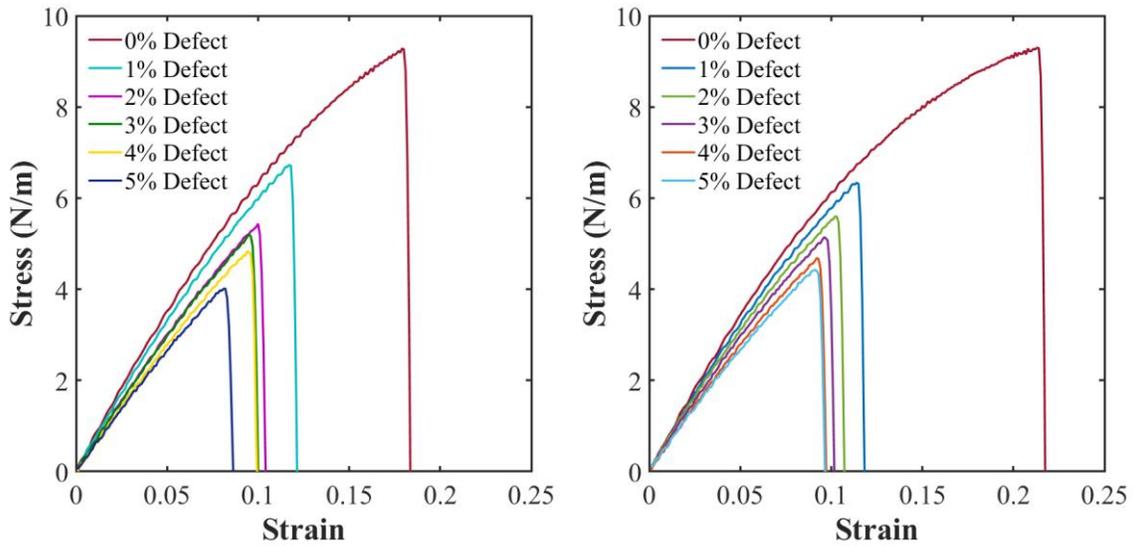

Figure 7: Stress-strain relationship for the defected $TiS_2$ samples altering the density of defect while applied strain along (a) armchair (b) zigzag direction

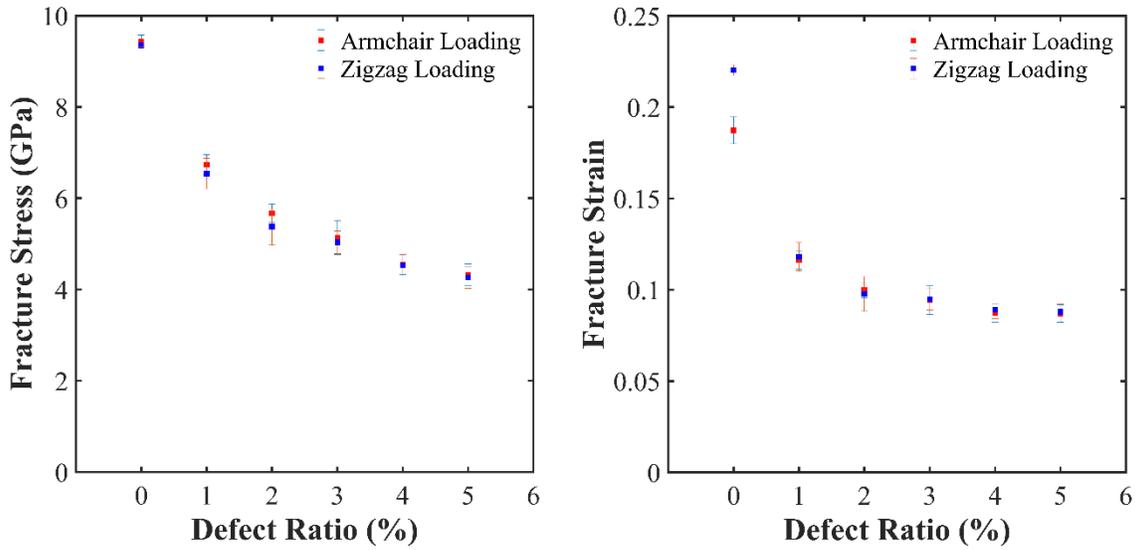

Figure 8: Change of fracture stress and strain with estimated error bar for the variation of defect density

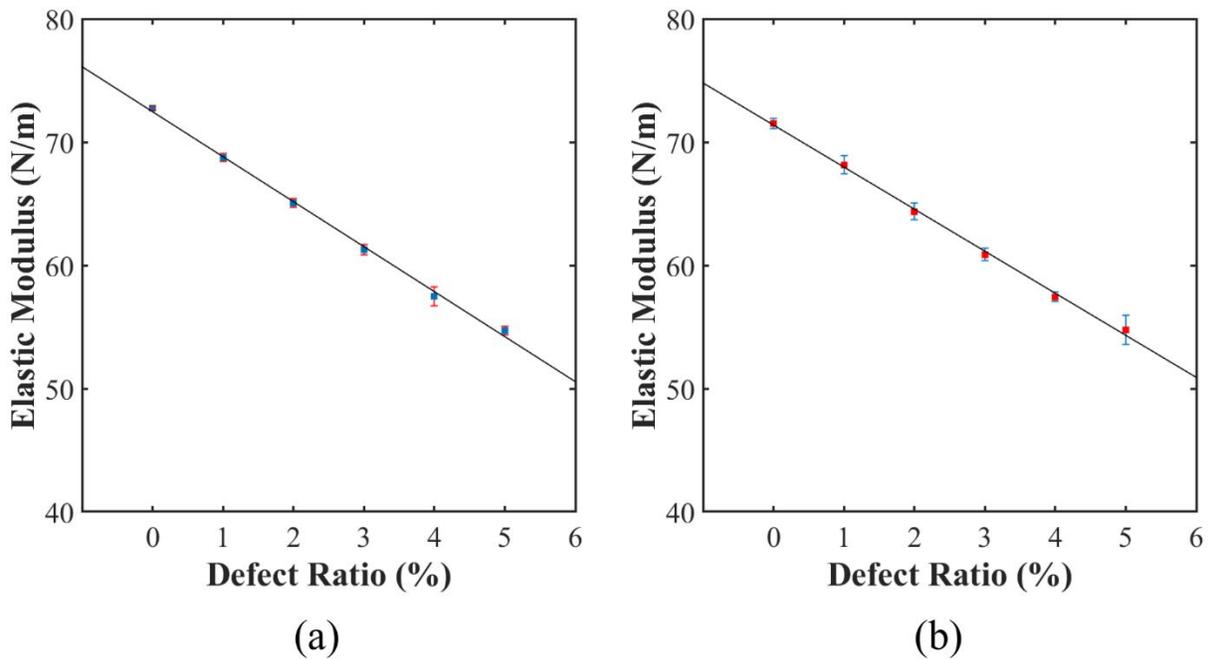

Figure 9: Change of elastic modulus (a) armchair (b) zigzag directional tension with estimated error bar for the variation of defect density

### 4.6 Fracture Mechanism

We observed fracture mechanism of pristine and defected $TiS_2$ nanosheet for both armchair and zigzag tension. As uniaxial tension is applied on 1T-$TiS_2$ sheet, stress begins to develop in the structure. However, stress does not develop uniformly everywhere rather in some distinct points. Nucleation starts from the places where the stress concentration is significantly higher. Since stress concentration around defects is always greater, defected points act as nucleation center for defected structure. For both the pristine and defected sheet, crack forms and propagates perpendicular to loading direction (zigzag for armchair loading, armchair for zigzag loading) initiating from the nucleation point. Crack propagation speed for armchair and zigzag crack is estimated to be around 6 km/s and 5 km/s respectively. Crack propagation speed is calculated by dividing the increase in crack length with respect to time. Although there are inclined bonds both in armchair and zigzag direction, in case of armchair direction, the inclined bonds are at ±60° whereas they are at ±30° for zigzag. Consequently, there are more branching in terms of zigzag loading than armchair loading during crack propagation. Again, when there are arbitrarily distributed point defects in the sheet, the crack path tends to follow an easier path along the defects for dissipating energy rather than breaking new bonds to facilitate branching. Therefore, there are less and almost no branching observed in zigzag and armchair loading condition respectively for a randomly distributed defected sheet.

(a)

(i)

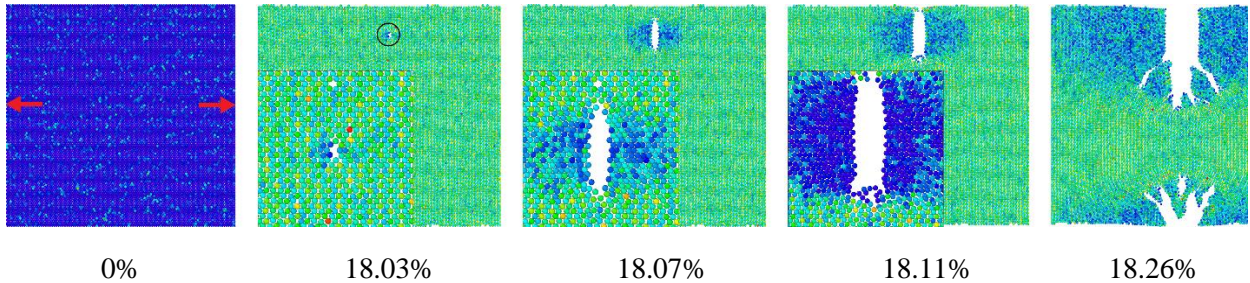

0%    18.03%    18.07%    18.11%    18.26%

(ii)

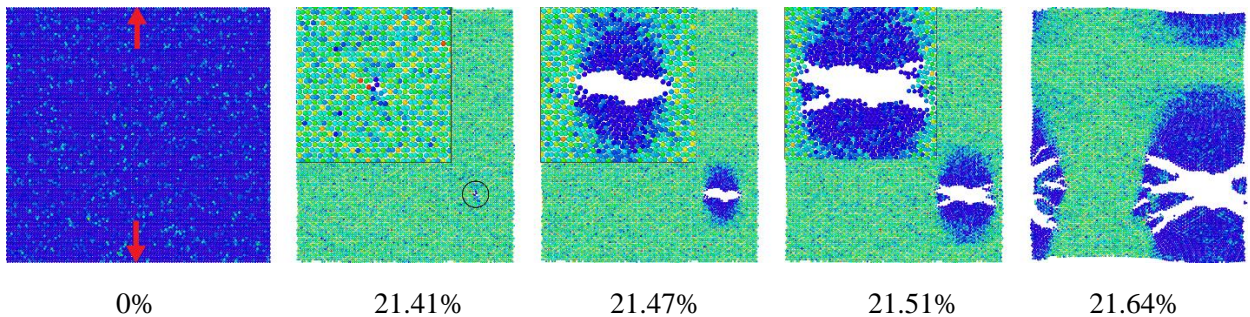

0%    21.41%    21.47%    21.51%    21.64%

(b)

(i)

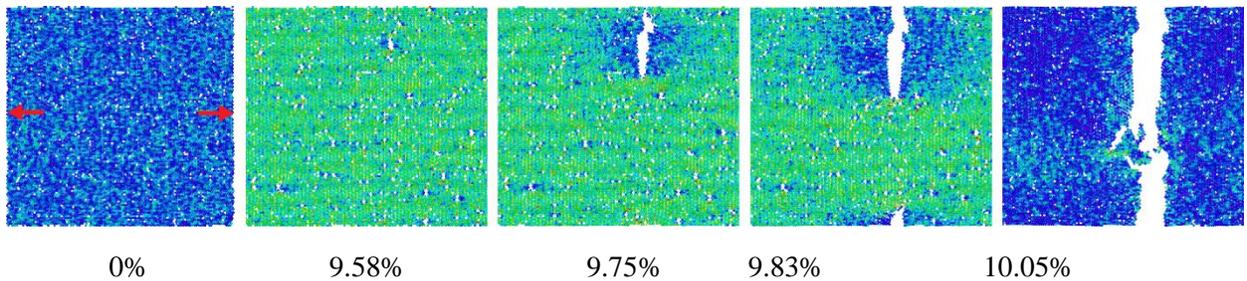

0%    9.58%    9.75%    9.83%    10.05%

(ii)

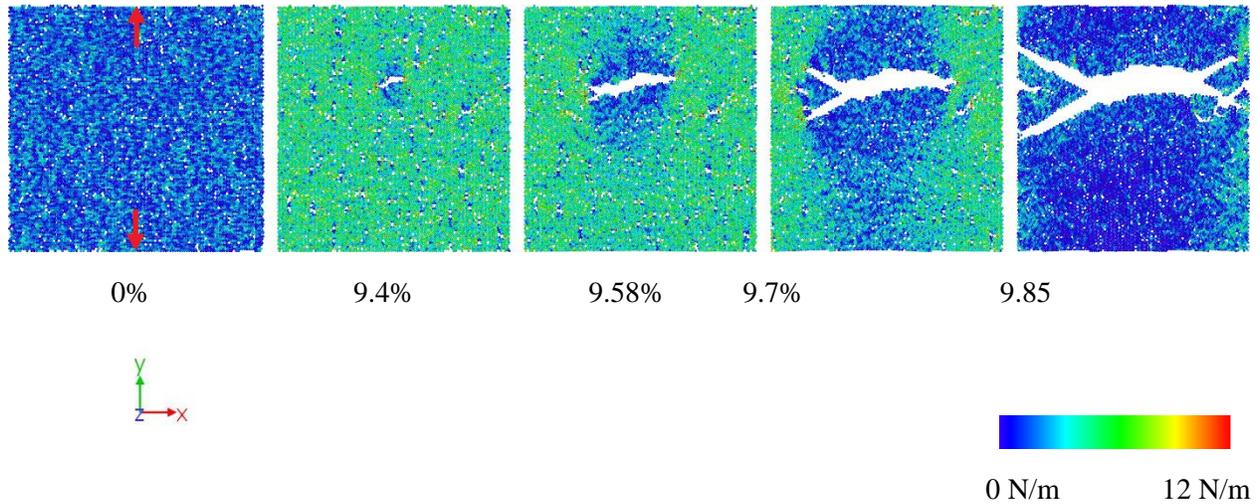

| 0% | 9.4% | 9.58% | 9.7% | 9.85 |

0 N/m      12 N/m

Figure 10: The stress distribution, deformation process and crack propagation of SLTiS$_2$ in (a) pristine (b) 3% defected structure, uniaxial tension along (i) armchair (ii) zigzag direction. The color bar shows the stress in N/m. Here, X and Y axis represents armchair and zigzag edges, respectively. Red colored arrow shows the straining direction.

## 5. CONCLUSIONS

For recapitulation, we carried out molecular dynamics simulations to inspect the mechanical properties and fracture behavior of pristine and defected TiS$_2$ structure. We checked the effect of temperature and strain rate on the pristine structure of the material. Temperature plays a significant role on the mechanical properties of the nanosheet. Increase in temperature leads to degradation of not only in fracture strength and strain but also in elastic modulus. On the other hand, increase in strain rate causes escalation in fracture strength. However, elastic modulus remains constant with the variation. We also studied the effect of different types of defects and defect ratio on the material. Defected samples always undermines the fracture strength of the pristine one. And increasing defect ratio causes decrease in material integrity and thus weakens the strength of the material. And finally, fracture mechanism reveals that branching is dominant in zigzag tension.

## 6. ACKNOWLEDGEMENT

The authors of this paper would like to acknowledge the technical support provided by the Multiscale Mechanical Modeling and Research Network (MMMRN) group of BUET. M.M.I acknowledges the support from Wayne State University startup funds.

# Supplementary Information

# Atomic Scale Insights Into The Mechanical Characteristics of Monolayer 1T-Titanium Disulphide: A Molecular Dynamics Study

Tanmay Sarkar Akash,[a] Rafsan A.S.I. Subad,[a] Pritom Bose,[a] and Md Mahbubul Islam*[b]

[a]Department of Mechanical Engineering, Bangladesh University of Engineering and Technology, Dhaka-1000, Bangladesh

[b]Department of Mechanical Engineering, Wayne State University, 5050 Anthony Wayne Drive, Detroit, MI- 48202, USA

*Corresponding Author Tel.: 313-577-3885; E-mail address: mahbub.islam@wayne.edu

**SUPPLEMENTARY FIGURES**

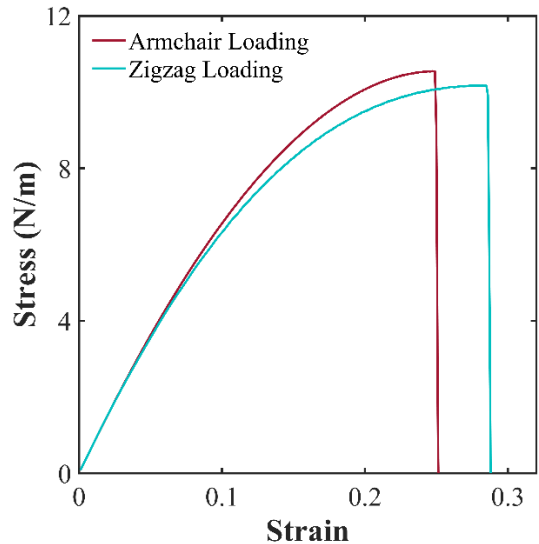

Figure S1: Stress-strain curve for pristine SLTiS$_2$ (temperature 10K, strain rate $10^{-9}$ s$^{-1}$)

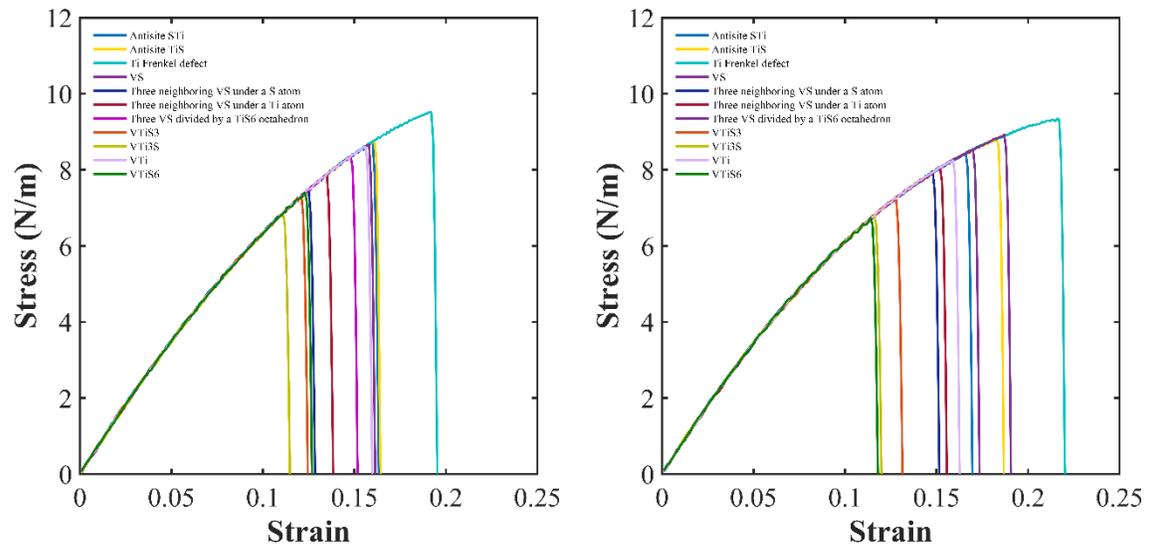

Figure S2: Stress-strain relationship for defected monolayer TiS$_2$ nanosheet (temperature 300K, strain rate $10^{-9}$ s$^{-1}$)

(a)

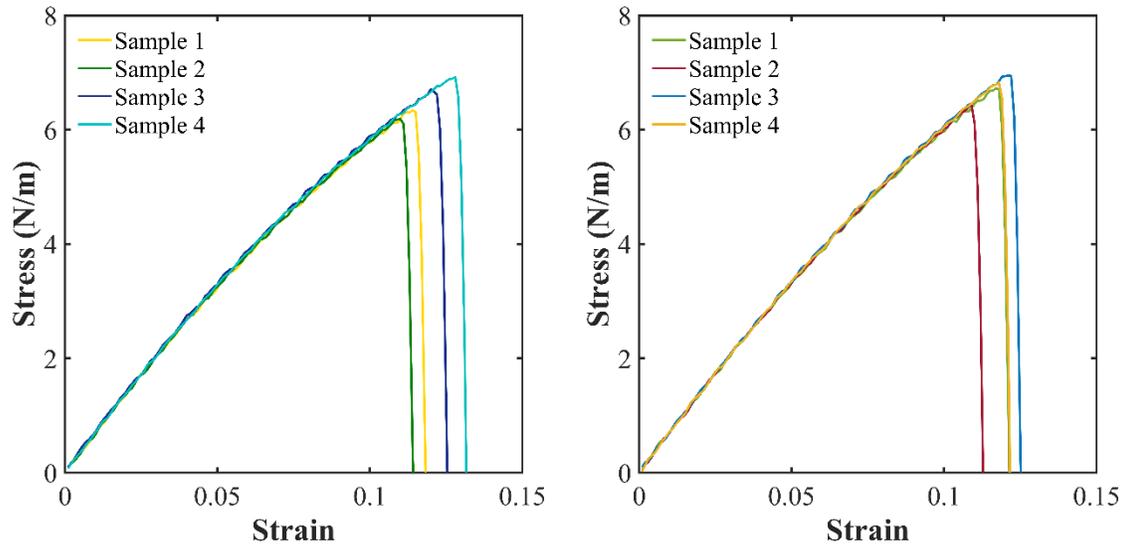

(i)  (ii)

(b)

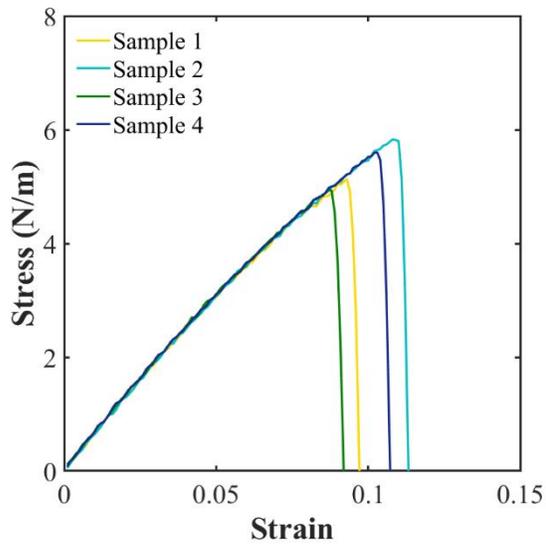

(i)

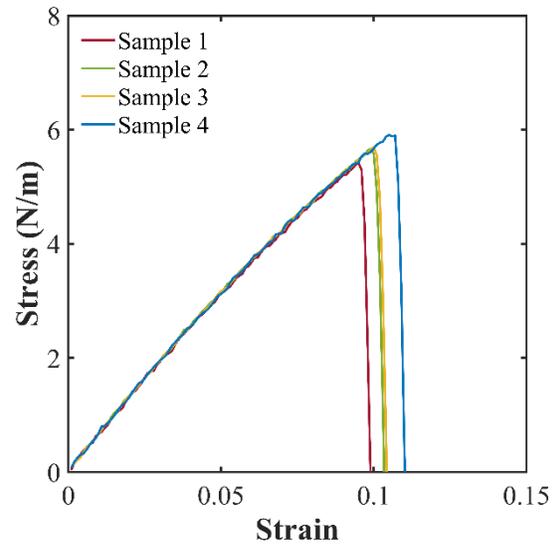

(ii)

(c)

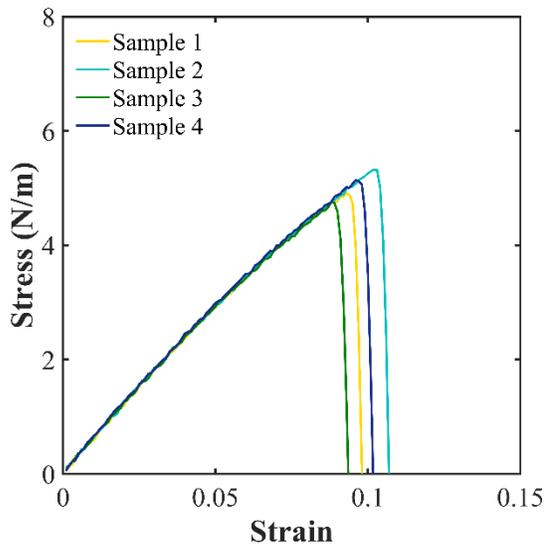

(i)

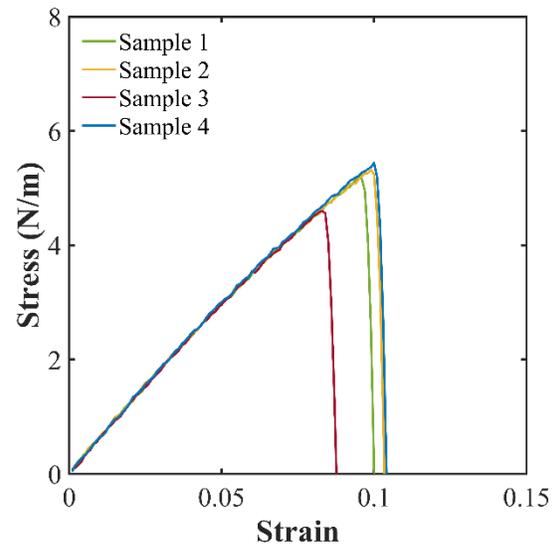

(ii)

(d)

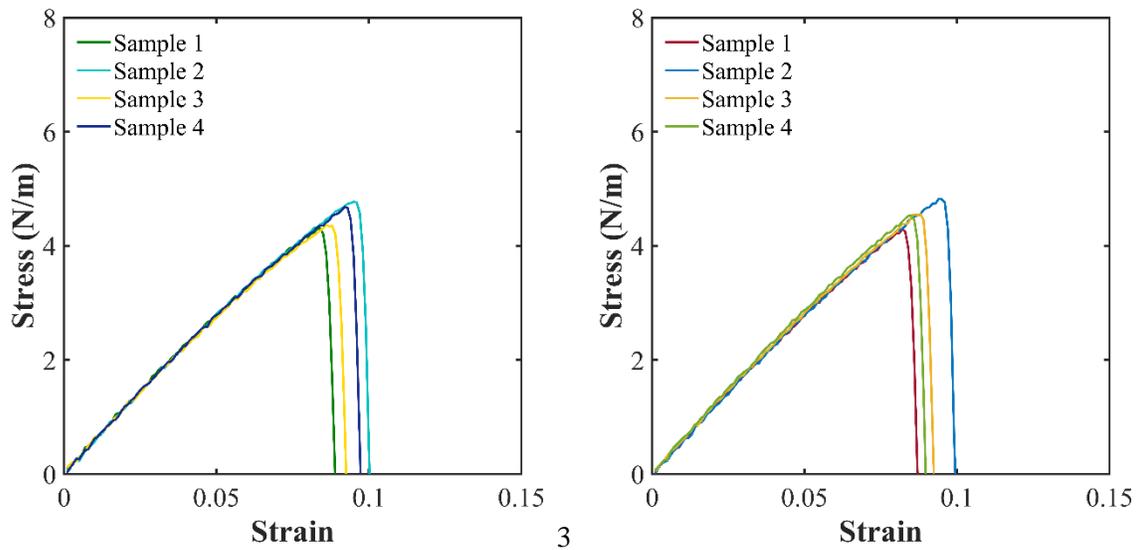

(i)                  (ii)

(e)

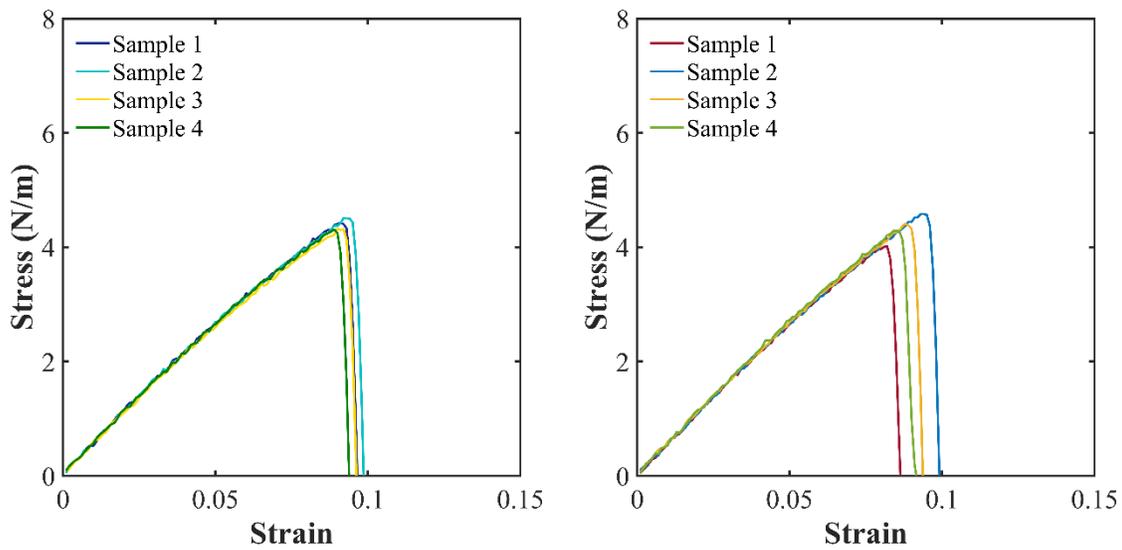

(i)                  (ii)

Figure S3: Statistical analysis for defect ratios, (a), (b), (c), (d), and (e) for 1%, 2%, 3%, 4%, 5% defect concentration and each (i), (ii) designates armchair and zigzag directional loading condition accordingly (temperature 300K, strain rate $10^{-9}$ s$^{-1}$)